\documentclass[aps,prl, twocolumn,reprint,amsmath,amssymb,superscriptaddress,nobibnotes,showpacs]{revtex4-1}

\usepackage{bbm} 
\usepackage{graphicx}
\usepackage{dcolumn}
\usepackage{bm}
\usepackage{amsmath}
\usepackage{amssymb,amsthm}
\usepackage{color}
\usepackage[sans]{dsfont}

\definecolor{nblue}{rgb}{0.2,0.2,0.7}
\definecolor{ngreen}{rgb}{0.2,0.6,0.2}
\definecolor{nred}{rgb}{0.7,0.2,0.2}
\definecolor{nblack}{rgb}{0,0,0}

\newcommand{\ket}[1]{|#1\rangle}
\newcommand{\braket}[2]{\langle{#1}|{#2}\rangle}

\newcommand{\fig}[1]{Fig.~\ref{#1}}
\newcommand{\tab}[1]{Table \ref{#1}}

\def\A{\mathcal{A}}

\def\beq{\begin{eqnarray}}
\def\eeq{\end{eqnarray}}
\def\bfig{\begin{figure}}
\def\efig{\end{figure}}

\newcommand{\abs}[1]{\left|{#1}\right|}

\begin{document}

\title{Non-monotonic quantum to classical transition in multiparticle interference}

\author{Young-Sik Ra}
\affiliation{Department of Physics, Pohang University of Science and Technology (POSTECH), Pohang, 790-784, Korea}

\author{Malte C. Tichy}
\affiliation{Physikalisches Institut der Albert-Ludwigs-Universit\"at, Hermann-Herder-Str.~3, D-79104 Freiburg}

\author{Hyang-Tag Lim}
\affiliation{Department of Physics, Pohang University of Science and Technology (POSTECH), Pohang, 790-784, Korea}

\author{Osung Kwon}
\affiliation{Department of Physics, Pohang University of Science and Technology (POSTECH), Pohang, 790-784, Korea}

\author{Florian Mintert}
\affiliation{Physikalisches Institut der Albert-Ludwigs-Universit\"at, Hermann-Herder-Str.~3, D-79104 Freiburg}
\affiliation{Freiburg Institute for Advanced Studies, Albert-Ludwigs-Universit\"at, Albertstrasse 19, D-79104 Freiburg}

\author{Andreas Buchleitner}
\affiliation{Physikalisches Institut der Albert-Ludwigs-Universit\"at, Hermann-Herder-Str.~3, D-79104 Freiburg}

\author{Yoon-Ho Kim}
\affiliation{Department of Physics, Pohang University of Science and Technology (POSTECH), Pohang, 790-784, Korea}

\date{\today}
\begin{abstract}
We experimentally demonstrate the non-monotonic dependence of genuine many-particle interference signals on the particles' mutual distinguishability. Our theoretical analysis shows that such non-monotonicity is a generic feature of the quantum to classical transition in multiparticle correlation functions of more than two particles.
\end{abstract}
\maketitle
 A quintessential ingredient of quantum physics is the  superposition principle. It becomes manifest in the (self-) interference of single particles, as observed for systems ranging from photons~\cite{ref:grangier} to fullerene molecules~\cite{arndt05}. These phenomena rely on the \emph{coherence} of the single-particle wave-function, which guaranties that the different pathways a single particle can take to a detector -- {\it e.g.}~through the left or through the right slit in a double-slit experiment -- remain indistinguishable.

 Interaction with the environment, however, may convey which-path information to the environment, and then inevitably leads to decoherence \cite{cohevol02}. Thereby, it jeopardizes the ideal interference pattern and induces the quantum-to-classical transition \cite{arndt91}: \emph{probabilities} instead of amplitudes need to be summed to obtain event probabilities. The stronger the decoherence, the weaker is the interference signal \cite{englert96}. Expectation values of observables therefore depend \emph{monotonically} on the strength of decoherence, and a \emph{monotonic} transition between quantum and classical expectation values takes place.

 The superposition principle also applies to \emph{many}-particle wave-functions, with entanglement \cite{ref:schrodinger} and many-particle interference \cite{Peruzzo:2011dq} as immediate consequences. For the observation of the latter, the particles' mutual indistinguishability is necessary. Given that, one observes, {\it e.g.}, the 
 Hong-Ou-Mandel (HOM) effect~\cite{ref:hong,ref:pittman}: When a single photon is incident on each of the two input modes of a balanced beam splitter, the {\em two-particle} Feynman-paths of ``both photons reflected'' and ``both photons transmitted'' interfere destructively, leading to the strict suppression of the event with one particle per output mode.
 
 Much as in the case of single-particle interference discussed above, the \emph{transition} between \emph{distinguishable} and \emph{indistinguishable} particles is monotonic also in the HOM setting: the interference-induced suppression of the balanced output event fades away monotonically with increasing particle distinguishability~\cite{ref:hong,al:2009vn}, since particle-distinguishability is tantamount of the distinguishability of many-particle paths. Therefore, conveying information on the interfering particles' identity provides which-path information in the space of many-particle paths. Monotonic distinguishability-dependence of bunching events (where all particles are detected in one output mode) was also observed for four \cite{ref:ou1} and six \cite{Niu:2009pr} photons, while events other than bunching were not considered yet for more than two photons.

 In our present contribution, we focus on many-particle interference effects in multi-port output events {\em distinct} from bunching. In contrast to the hitherto established scenario sketched above, we will see that the distinguishibility-induced suppression of multi-particle interference signals is, in general, a {\em non-monotonic} function of the particles' distinguishability, provided that more than two particles are brought to interference \cite{ref:malte}. This is explained by a hierarchy of different orders of many-particle interferences which dominate the total interference signal at different stages of the distinguishability transition. Only when reduced indistinguishability defines two {\em unambiguous} alternatives does one enter the monotonic realm of the quantum to classical transition.

\bfig[t]
\includegraphics[width=\linewidth]{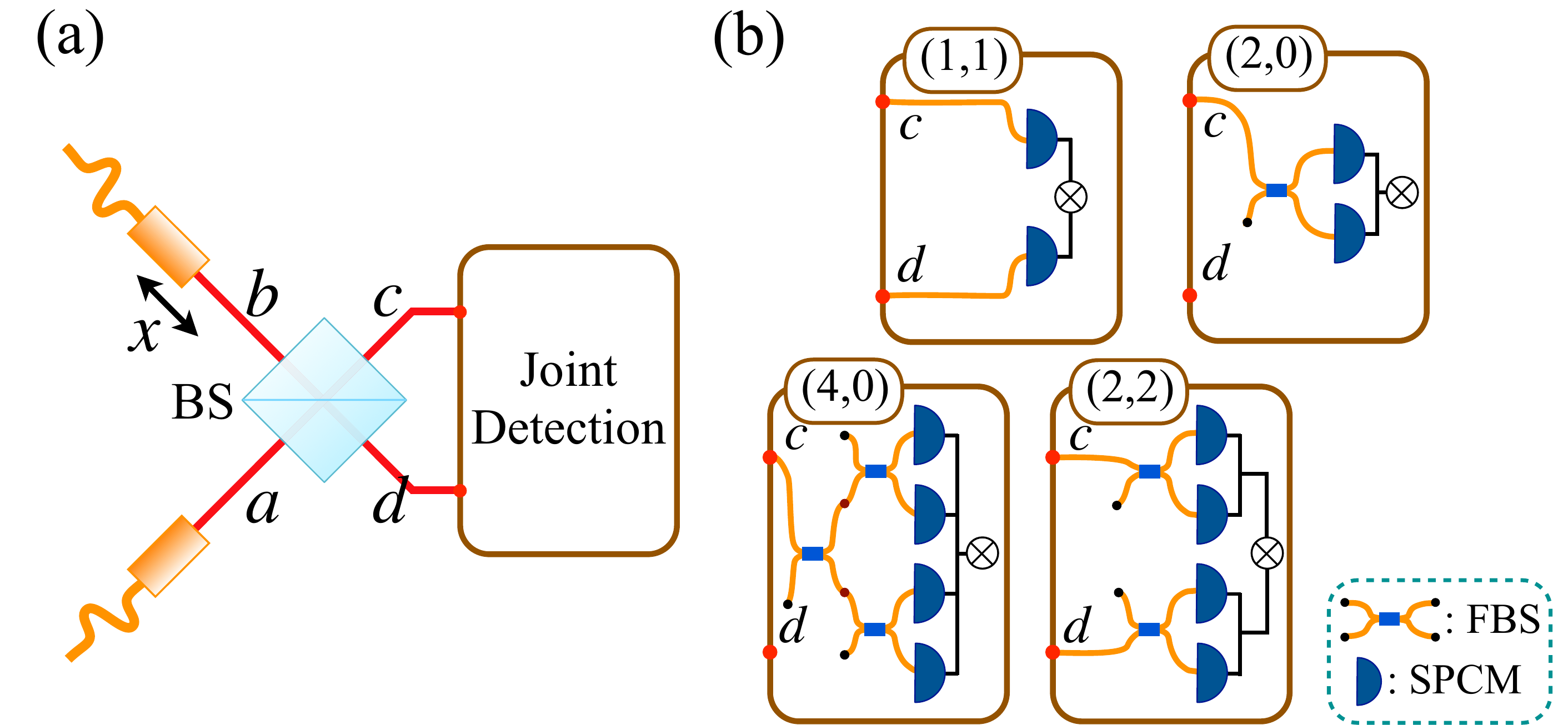}
\caption{(a) HOM type interferometer. The path delay $x$ controls the distinguishability between the photons from mode $a$ and $b$ which then scatter on the 50:50 beam splitter (BS). (b) Joint detection of (\textit{m,n}) photons at the output modes $c$ and $d$, which is realized by concatenating fiber beam splitters (FBSs, 50:50) and single-photon counting modules (SPCMs).} \label{fig:scheme} \efig 

 Let us first describe our experimental setup shown in \fig{fig:scheme}(a). Single-photon or two-photon Fock-states are injected into each input mode ($a$,$b$), and the particle distributions at the output modes $c$ and $d$ of the beam splitter (BS, 50:50) are measured while changing the path delay $x$ of the mode $b$, thereby varying the distinguishability of the photons in the input modes. The single- and two-photon Fock states are generated via the non-collinear frequency-degenerate spontaneous parametric down-conversion (SPDC) process at a 2 mm-thick $\beta$-barium-borate (BBO) crystal pumped by a femtosecond laser pulse (central wavelength: 390 nm, average power: 120 mW) which is focused onto the BBO by a lens of 300 mm focal length. The generated photons are centered at 780 nm and filtered by interference filters of 4 nm full width at half maximum (FWHM). The photons are coupled into two single-mode fibers located at a distance of 430 mm from the BBO, where each fiber corresponds to an input mode ($a$, $b$). Two-fold counts by simultaneous single-photon detection between the two fibers were recorded at a rate of 13 kHz, by combination of single-photon counting modules (SPCMs, Perkin-Elmer SPCM-AQ4C)  and a coincidence circuit with 8 ns coincidence window. At the fibers' output, half-wave plates and quarter-wave plates adjust the photons' polarizations, and a motorized stage connected to the fiber tip at the mode $b$ changes $x$. The quantum state of the input modes is proportional to $\sum_{j=0}^{\infty} \eta^j \ket{j,j}_{a,b} $, where $|\eta|^2$ is the probability for photon pair generation, and coincidence detection of two or four photons in the output modes projects the generated state onto the initial state $\ket{1,1}_{ab}$ or $\ket{2,2}_{ab}$, respectively. The photon distribution at the output modes is measured by combining 50:50 fiber beam splitters (FBSs, 2$\times$2 single mode coupler) and SPCMs, as depicted in \fig{fig:scheme}(b), such that $m$ and $n$ particles in the first and second output mode, respectively, are detected simultaneously, to define a $(m,n)$-event. The distinguishability of the photons in modes $a$ and $b$ depends on $x$. When $x$ = 0 $\mu$m, the photons are perfectly indistinguishable, and exhibit ideal interference. When $x$ is much larger than the single-photon coherence length $l_c =$134 $\mu$m, the particles can be treated as distinguishable, so many-particle interference does not occur. 
 
 In a first step, we study the effect of the distinguishability transition of two photons. While changing $x$, we measured the probabilities of the two-photon events (1,1) and (2,0). The resulting data in \fig{fig:data}(a) show typical two-photon HOM-type interferences~\cite{ref:hong,ref:yhkim}. As expected~\cite{ref:hong}, the event probabilities are complementary and depend monotonically on the photons' distinguishability.
\bfig[t]
\includegraphics[width=3.2in]{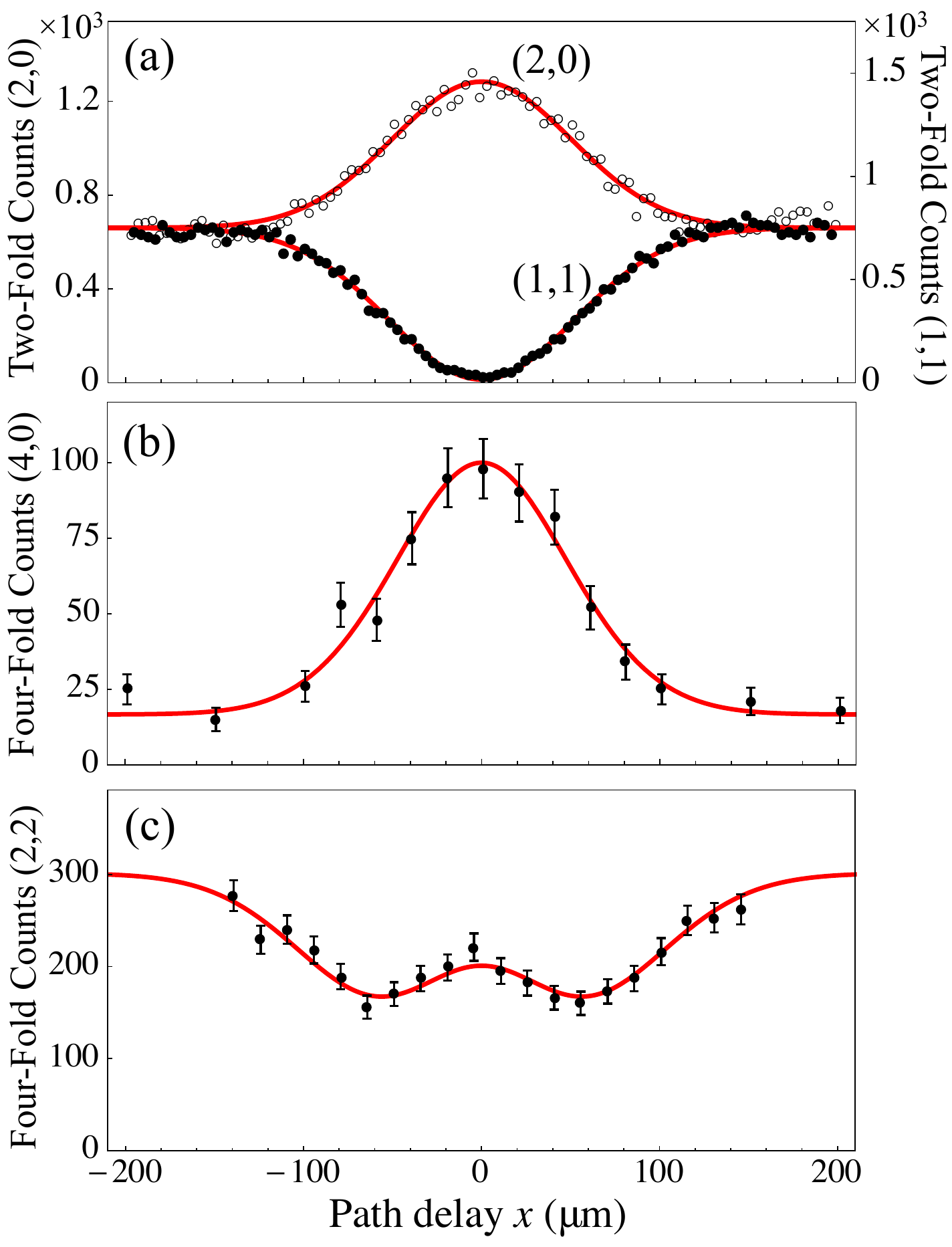}
\caption{
Event probabilities (in absolute coincidence counts) {\it vs.} path delay $x$. (a) HOM-type interference of two-photons, with a clear enhancement of the (2,0) event (two photons in one output mode; open circles), and the corresponding suppression of the (1,1) event (one photon in each output mode; filled circles). Errors are below 36 counts per bin, and omitted for clarity. Four photon interference with two photons per input mode is monitored in (b) by the (4,0) (integration over $14100\ {\mathrm s}$), and in (c) by the (2,2) (integration over $9600\ {\mathrm s}$) event probabilities, respectively. The red lines are the event probabilities as predicted by (5) -- within the statistical error in perfect agreement with the experimental data.}
\label{fig:data} \efig 
 Next, the four-photon events, (4,0) and (2,2), are recorded, and we monitor the resulting four-photon distinguishability transition. A clear enhancement of (4,0)-events for indistinguishable particles is observed in \fig{fig:data}(b) and, indeed, this enhancement fades away monotonically as the particles are tuned to become distinguishable by increasing $x$. The event (2,2), however, in \fig{fig:data}(c), exhibits minima symmetrically displaced from the origin $x=0$, and the distinguishability transition, as induced by continuous variation of $x$ from $0$ to $x\gg \ell_c$, is not any longer unambiguously reflected by a monotonic $x$-dependence of the many-particle interference signal. This is indicative of an intricate interplay of many-particle interference and distinguishability transition, which we will elucidate in the following. 
 
 In our setup sketched in \fig{fig:scheme}(a), the path delay $x$ controls the (in-)distinguishability between photons injected into the modes $a$ and $b$, respectively, by controlling their arrival times at the BS. A photon with arrival time $t_i$ is described as
\beq
\A_{t_i}^\dagger \ket{0}=\ket{1_{t_i}}=
\int_{-\infty}^{\infty} \mbox{d} \omega \frac{1}{\sqrt \pi
\Delta \omega} e^{- \frac{(\omega-\omega_0)^2}{2 \Delta\omega^2}
} e^{i \omega t_i} \A_{\omega}^\dagger \ket{0},
\eeq
 where $\omega_0$ is the central frequency, $\Delta \omega$ is the frequency width, and $\A^{\dagger}_{\omega}$ creates a photon of frequency $\omega$. The distinguishability of two photons with different arrival times $t_1$ and $t_2$ is quantified by 
\begin{equation}
\abs{\alpha}^2 = \mathrm{exp}\left({-\Delta\omega^2 {(t_2-t_1)}^2 /2} \right)\, ,\, t_2-t_1=\frac{x}{c}\, ,
\end{equation}
 with $\alpha=\braket{1_{t_1}}{1_{t_2}}$. The photons are strictly indistinguishable for $\abs{\alpha}^2=1$, and fully distinguishable for $\abs{\alpha}^2=0$. Consequently, the $t_2$  photon can be decomposed -- with respect to its support on the time axis -- by orthogonal projection onto one strictly indistinguishable component $\alpha \ket{1_{t_1}}$, and onto one (orthogonal) fully distinguishable component $\sqrt{1-\alpha^2} \ket{\widetilde{1}_{t_1}}$ \footnote{Since the phase of $\alpha$ is not observable in our setting, we assume $\alpha\in \mathbbm{R}$ in the following, without loss of generality.}.  A two-photon state with one photon in each input modes of arrival time $t_1$ and $t_2$ thus reads
\beq
a_{t_1}^{\dagger} b_{t_2}^{\dagger} \ket{0}&=& \alpha \ket{1,1}_{ab}+\sqrt{1-\alpha^2} \ket{1,\widetilde{1}}_{ab} \label{eq:twophotdec}
\eeq
after substitution of the specific input modes $a$, $b$ for $\A$. Note that we have omitted the subscript $t_1$ for simplifying the notation. Herein, the components $\ket{1,1}_{ab}$ and $\ket{1,\widetilde{1}}_{ab}$ are orthogonal, with strictly indistinguishable particles in $\ket{1,1}_{ab}$, and fully distinguishable particles in $\ket{1,\widetilde{1}}_{ab}$. The continuous parametrization of the distinguishability transition by $x$ is now replaced by continuous parametrization by $\abs{\alpha}^2\in [ 0;1]$. Only the indistinguishable component in (3) defines indistinguishable two-particle paths which can interfere, and the resulting interference signal therefore has to fade away monotonically with decreasing $\abs{\alpha}^2$ -- this is just the HOM scenario, as reproduced by our experimental results in \fig{fig:data}(a).

 Something {\em qualitatively} new happens when more than one particle is injected in each input mode: The orthogonal decomposition analogous to (3) gives rise 
to {\em cross terms} between the distinguishable and indistinguishable components. Specifically, for two photons injected into each mode, we obtain 
\beq
\frac{1}{2} \left(a_{t_1}^{\dagger} \right)^2 \left(b_{t_2}^{\dagger}\right)^2 \ket{0}&=& \alpha^2 \ket{2,2}_{ab}+\sqrt{2}\alpha\sqrt{1-\alpha^2} \ket{2,1\widetilde{1}}_{ab} \nonumber \\
&& +(1-\alpha^2) \ket{2,\widetilde{2}}_{ab}\, , \label{eq:decomp4}
\eeq
with the cross term $\propto \ket{2,1\widetilde{1}}_{ab}$ mutually orthogonal to $\ket{2,2}_{ab}$ and $\ket{2,\widetilde{2}}_{ab}$. In precise analogy to the HOM scenario,
 the $\ket{2,2}_{ab}$ component defines indistinguishable four-particle paths that give rise to an interference signal with amplitude $\alpha^2$, fading away with increasing amplitude $(1-\alpha^2)$ of the (distinguishable) $\ket{2,\widetilde{2}}_{ab}$ component. In contrast, the novel term $\propto \ket{2,1\widetilde{1}}_{ab}$ represents relative timing such that one of the two particles in mode $b$ is strictly indistinguishable from those in mode $a$, while the other one is fully distinguishable. Therefore, this term gives rise to a three-particle interference signal through coherent superposition of indistinguishable three-particle paths, with amplitude $\sqrt{2}\alpha\sqrt{1-\alpha^2}$. The probability to detect $m$ particles in output mode $c$ and $n$ particles in output mode $d$, given $N=m+n$ particles on input, is then generally given by
\beq
P^{(N;m,n)}(x) =\sum_{\mathrm{type}} p_{\mathrm{type}}^{(N;m,n)}\ W_{\mathrm{type}}^{(N)}(x)\ , \label{eq:totalprobability}
\eeq
where we need to sum over the various {\em distinguishibility types} of the contributions as they emerge in (4): strictly indistinguishable (indis), partially distinguishable (inter), and fully distinguishable (dist). The associated detection probabilities $p_{\mathrm{type}}^{(N;m,n)}$ are determined by the geometry of the experimental set-up, computed by mapping input on output modes via
\beq
a^\dagger \rightarrow \frac{1}{\sqrt 2}\left(c^\dagger+ d^\dagger\right) , \
b^\dagger\rightarrow \frac{1}{\sqrt 2}\left(c^\dagger- d^\dagger \right) \, , \label{eq:evolu}
\eeq
and listed in \tab{ptable} for the specific events experimentally probed in \fig{fig:data}.
\bfig
\includegraphics[width=\linewidth]{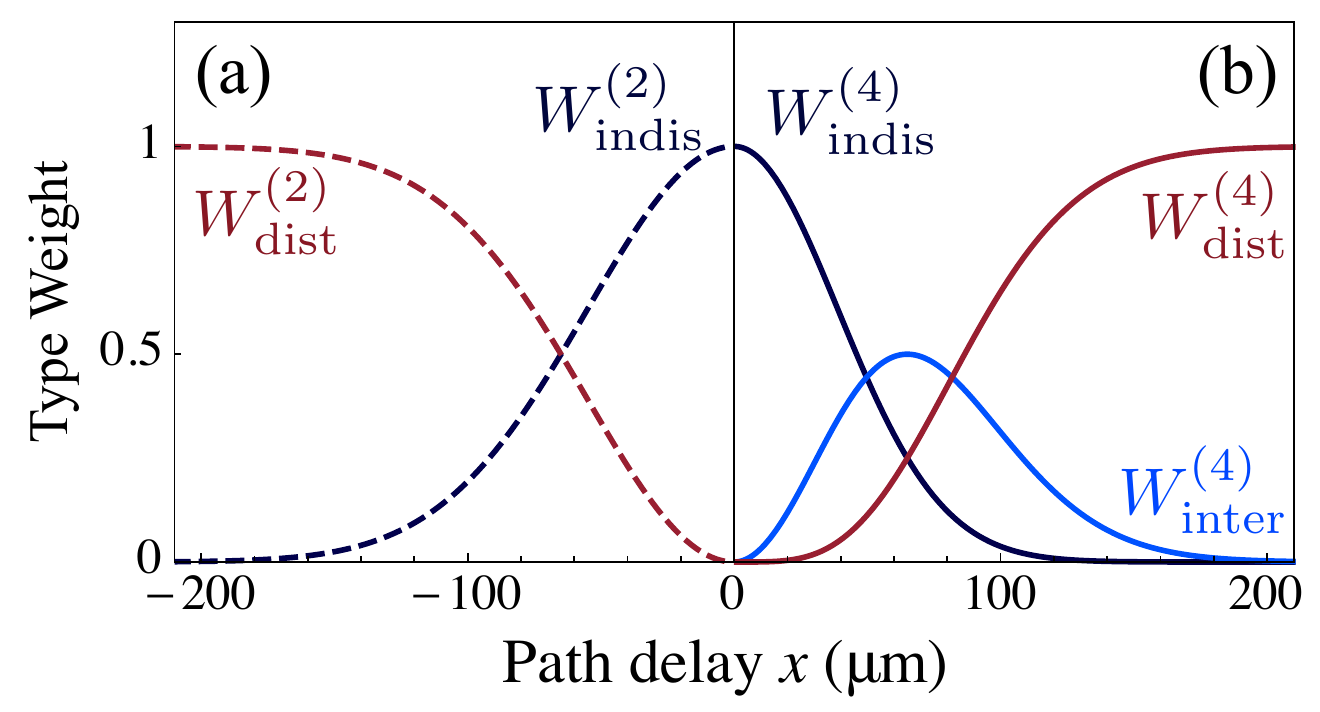}
\caption{
Weights of the contributions of strictly indistinguishable ($W^{(N)}_{\mathrm{indis}}$), of partially distinguishable ($W^{(N)}_{\mathrm{inter}}$), and of fully distinguishable ($W^{(N)}_{\mathrm{dist}}$) particles to the many-particle wave-function, as a function of the path delay $x$ which continuously parametrizes the particles' distinguishability. (a) For $N=2$, partial distinguishability does not arise, and $W^{(N)}_{\mathrm{indis}}$ and $W^{(N)}_{\mathrm{dist}}$ are monotonic functions of $x$, leading to a monotonic distinguishability transition in $P^{(2;m,n)}(x)$. (b) In contrast, for $N=4$, $W^{(N)}_{\mathrm{inter}}$ contributes non-monotonically to the event probability, and induces the non-monotonic behavior observed in \fig{fig:data}(c).}
\label{fig:distinguishability} \efig 
 \begin{center}
\begin{table}[t]
\begin{tabular}{c|rr}
$p^{(2;m,n)}$&~~indis&~~dist\\
\hline
$~(2,0)~$&1/2&1/4\\
$~(1,1)~$&0&1/2
\end{tabular}~~~~
\begin{tabular}{c|rrr}
$p^{(4;m,n)}$&~~indis&~~inter&~~dist\\
\hline
$~(4,0)~$&3/8&3/16&1/16\\
$~(2,2)~$&1/4&1/8&3/8
\end{tabular}\caption{
Detection probabilities $p^{(N;m,n)}$ of $m$ ($n$) particles in mode $c$ ($d$) as derived from the mutually orthogonal amplitudes of the strictly indistinguishable (indis), partially distinguishable (inter) and fully distinguishable (dist) contributions in (3,4), for $N=2$ (left panel) and $N=4$ (right panel). 
}\label{ptable}
\end{table} 
\end{center}

 The weights $W_{\mathrm{type}}^{(4)}(x)$ in (\ref{eq:totalprobability}) are simply given by squaring the amplitudes of the respective components in the orthogonal decomposition (4), where we baptize the cross term $\propto \ket{2,1\widetilde{1}}_{ab}$ by the label ``inter''. Their functional dependence on $x$ (derived from their dependence on $\alpha$ via (2)) is shown in \fig{fig:distinguishability}: while the weight of all particles being either strictly indistinguishable, $W_{\mathrm{indis}}^{(4)}(x)$, or fully distinguishable, $W_{\mathrm{dist}}^{(4)}(x)$, always depends monotonically on $x$, the weight $W_{\mathrm{inter}}^{(4)}(x)$ representing cross terms between distinguishable and indistinguishable components exhibits a non-monotonic $x$-dependence! Consequently, in the $N=4$ particle case studied in our experiment, three-particle interference contributions kick in while four-particle interference contributions fade away, thus giving rise to the non-monotonicity observed in \fig{fig:data}(c). The continuous lines in \fig{fig:data}(b,c) are derived from the $W_{\mathrm{type}}^{(4)}(x)$ in \fig{fig:distinguishability}, together with the detection probabilities from \tab{ptable} inserted in (\ref{eq:totalprobability}), and fit the experimental data perfectly well, without adjustable parameters. Note that bunching signals, as the $(4,0)$ event observed here, always exhibit a monotonic transition, as a consequence of the bosonic enhancement \cite{tichy10} of the associated, {\em unique} many-particle path, which induces the strict hierarchy $p^{(4;m,n)}_{\mathrm{indis}}>p^{(4;m,n)}_{\mathrm{inter}}>p^{(4;m,n)}_{\mathrm{dist}}$ in \tab{ptable}.

 In general, for $N/2$ particles per input mode there are $N/2-1$ cross terms of distinguishable and indistinguishable components (represented by the term $\propto W_{\mathrm{inter}}^{(N)}(x)$ in (\ref{eq:totalprobability})), which all depend non-monotonically on $x$, and this non-monotonicity is generically inherited by the derived event probabilities $P^{(N;m,n)}(x)$ (provided that it not be counterbalanced by the actual values of the detection probabilities $p_{\mathrm{indis/inter/dist}}^{(N;m,n)}$). \fig{fig:6particles} shows an example for $N=6$.
\bfig
\includegraphics[width=\linewidth]{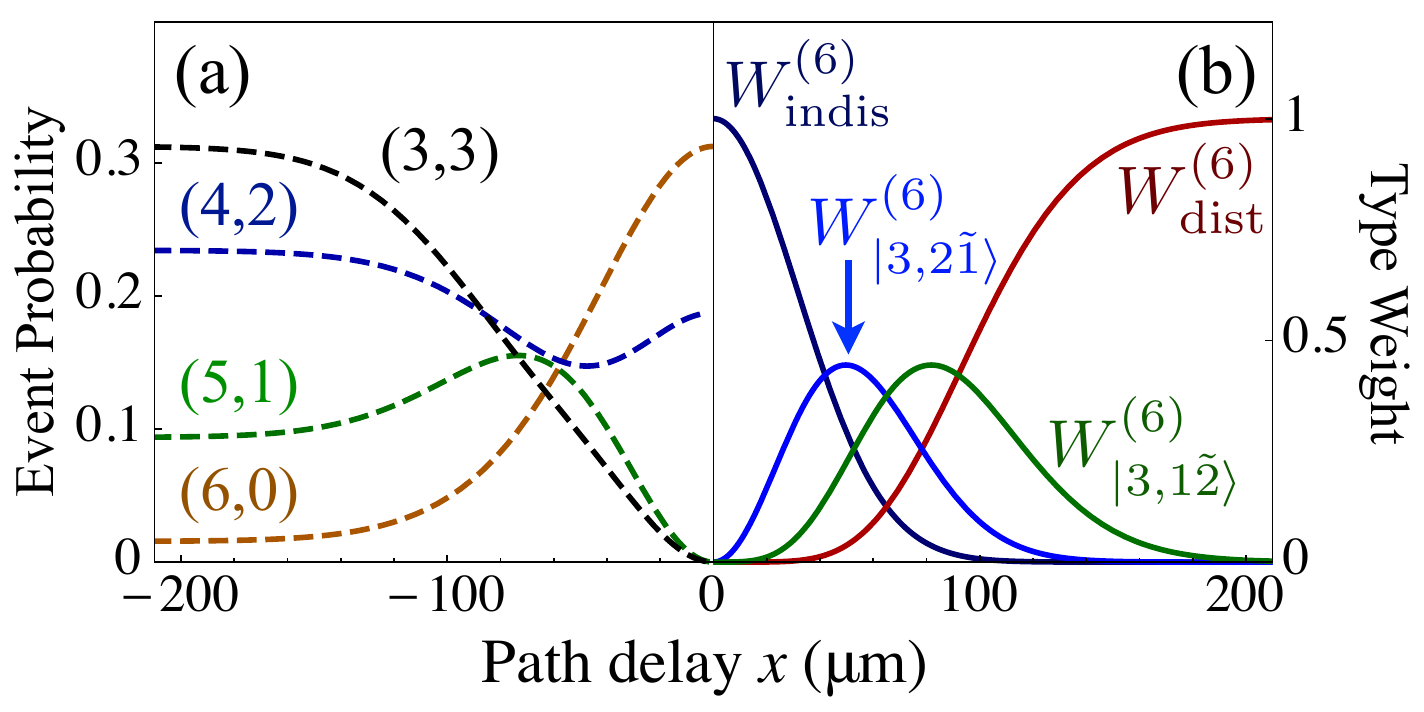}
\caption{ 
(a) Event probabilities $P^{(N;m,n)}(x)$, for $N=m+n=6$, and $m=3,\ldots, 6$, for each three photons injected into modes $a$ and $b$ (see \fig{fig:scheme}(a)). (b) shows the associated weights of strictly indistinguishable, partially and fully distinguishable contributions $W^{(N)}_{\mathrm{indis/inter/dist}}$, with a notation strictly analogous to that introduced in (4). Much as in \fig{fig:distinguishability}(b), the partially distinguishable contributions' weights are non-monotonic in $x$.
}\label{fig:6particles} 
\efig
 Let us finally discuss how this generic non-monotonicity of many-particle interference signals in correlation functions such as $P^{(N;m,n)}(x)$ with respect to $x$ matches our intuition of the monotonicity of the quantum to classical transition under decoherence: To begin with, the latter is generally observed in the interference of indistinguishable single-particle paths at {\em one} detector, while genuine many-particle interference manifests in multiparticle correlation functions not accessible in single-particle experiments. On that level, the distinguishability transition -- possibly caused by environment interaction, {\it i.e.}~coupling to additional degrees of freedom -- reduces the order of many-particle interferences, {\it e.g.}~from six- to five- and four-particle, or from four- to three-particle interference, in the examples considered in Figs.~2-4. As a consequence of the non-monotonic dependence of the different orders of the many-particle interference contributions on the particles' mutual distinguishability, the transition from fully multi-particle multi-path to single-particle coherence is, in general, non-monotonic and, yet, fully consistent with our understanding of single particle decoherence and the quantum to classical transition. However, we speculate that different types of system-environment interaction ({\it e.g.}~two {\it vs.}~three- or many-particle interactions) may allow to distinguish different orders of many-particle interference, and that, reciprocally, the dependence of the various interference terms in (\ref{eq:totalprobability}) on the environment coupling parameters may serve as an analytic tool to distinguish distinct environment coupling mechanisms.

 Financial support by the National Research Foundation of Korea (2009-0070668 and 2009-0084473), through a DAAD/GenKo grant 50739824, and by the German National Academic Foundation (M.C.T.) is gratefully acknowledged.


\end{document}